\documentclass{ar2e}
\begin{document}
\input epsf.tex    

\input psfig.sty

\def\beq{\begin{equation}}
\def\eeq{\end{equation}}

\def\d{{\rm d}}

\jname{Annu. Rev. Nuc. Part. Sci.}
\jyear{2009}
\jvol{59}

\title{Advances in Inflation in String Theory}

\markboth{Baumann and McAllister}{Advances in Inflation in String Theory}

\author{Daniel Baumann
\affiliation{Department of Physics, Harvard University,
Cambridge, MA 02138}
Liam McAllister
\affiliation{Department of Physics, Cornell University,
Ithaca, NY 14853}}

\begin{abstract}
We provide a pedagogical overview of inflation in string theory.
Our theme is the sensitivity of inflation to Planck-scale physics, which we argue provides both the primary motivation and the central theoretical challenge for the subject.
We illustrate these issues through two case studies of
inflationary scenarios in string theory: warped D-brane inflation and axion monodromy inflation. Finally, we indicate how future observations can test scenarios of inflation in string theory.
\vspace{0.5cm}
\end{abstract}

\pagestyle{plain}
\maketitle

\newpage
\section{Introduction}

Recent advances in observational cosmology have brought us closer to a fundamental understanding of the origin of structure in the universe.
Observations of variations in the cosmic microwave background (CMB) temperature and of the spatial distribution of galaxies in the sky have yielded a consistent picture in which gravitational instability drives primordial fluctuations to condense into large-scale structures, such as our own galaxy.
Moreover, quantum field theory and general relativity provide an elegant microphysical mechanism, {{\it inflation}}, for generating these primordial perturbations during an early period of accelerated expansion.
The classical dynamics of this inflationary era \cite{Guth:1980zm, Linde:1981mu, Albrecht:1982wi}
explains the large-scale homogeneity, isotropy and flatness of the universe, while
quantum fluctuations during inflation lead to small inhomogeneities.  The general properties of the spectrum of inflationary inhomogeneities were predicted long ago \cite{Mukhanov:1981xt, Mukhanov:1985rz, Hawking:1982cz, Starobinsky:1982ee, Guth:1982ec, Bardeen:1983qw} and are in beautiful agreement with recent CMB observations, e.g.~by the Wilkinson Microwave Anisotropy Probe (WMAP)~\cite{Komatsu:2008hk}.

Although inflation is remarkably successful as a phenomenological model for the dynamics of the very early universe, a detailed understanding of the physical origin of the inflationary expansion has remained elusive.  In this review we will highlight specific aspects of inflation that depend sensitively on the ultraviolet (UV)  completion of quantum field theory and gravity, i.e.~on the field content and interactions at energies approaching the Planck scale.  Such issues are most naturally addressed in a theory of Planck-scale physics, for which string theory is the best-developed candidate.  This motivates understanding the physics of inflation in string theory.

\subsection{Inflation}

Inflation may be defined as a period of exponential expansion of space,
\begin{equation}
{\rm d} s^2 = - {\rm d} t^2 + e^{2 H t} {\rm d} {\bf x}^2\, , \qquad {\rm where} \quad \epsilon \equiv - \frac{\dot H}{H^2} < 1\, ,
\end{equation}
which arises if the universe is dominated by a form of stress-energy that sources a nearly-constant Hubble parameter $H$.
This requirement for accelerated expansion can be fulfilled by a range of qualitatively different mechanisms with varied theoretical motivations \cite{ Lyth:1998xn, Baumann:2008aq, BaumannTASI}.
For concreteness, we restrict ourselves to the simple case of single-field slow-roll inflation, where the inflationary dynamics is  described by a single order parameter $\phi$ (a fundamental scalar field or a composite field) with canonical kinetic term $\frac{1}{2} (\partial_\mu \phi)^2$ and potential energy density $V(\phi)$.
Prolonged accelerated expansion
then occurs if the {\it slow-roll} parameters are small
\begin{equation}
\label{equ:SR}
\epsilon \simeq \frac{M_{\rm pl}^2}{2} \left(\frac{V'}{V}\right)^2 \ll 1\, , \qquad |\eta| = M_{\rm pl}^2 \left|\frac{V''}{V}\right| \ll 1\, ,
\end{equation}
where the primes denote derivatives with respect to the inflaton field $\phi$.

In addition to smoothing the universe on large scales, inflation stretches quantum fluctuations of light
degrees of freedom ($m \ll H$), creating a spectrum of small
perturbations in the observed CMB temperature and polarization.  These perturbations are the key to structure formation and to tests of inflation, so we pause to explain them; see e.g.~Ref.~\cite{Baumann:2008aq, BaumannTASI} for a more detailed discussion.

In inflationary scenarios with a single inflaton field, the light degrees of freedom during inflation are the inflaton itself and the two polarization modes of the graviton.  Fluctuations of the inflaton lead to perturbations of the time at which inflation ends, and hence source perturbations in the energy density after inflation.
These fluctuations are visible to us as anisotropies in the CMB temperature.
Both the inflaton ({\it scalar}) fluctuations and the graviton ({\it tensor}) fluctuations source polarization
of the CMB photons.
Ref.~\cite{Zaldarriaga:1996xe, Kamionkowski:1996ks} made the important observation that
the spin-2 polarization field of the CMB photons may be decomposed into two distinct scalar or spin-0 modes:
\begin{itemize}
\item {\it $E$-mode}: this curl-free mode is characterized by polarization vectors that are radial around cold spots and tangential around hot spots on the sky, and is generated by both scalar and tensor perturbations.
    $E$-mode polarization and its cross-correlations with the CMB temperature fluctuations were first detected by DASI \cite{DASI} and have recently been mapped out in greater detail by WMAP \cite{Kogut:2003et}.
\item {\it $B$-mode}: this divergence-free mode is characterized by polarization vectors with vorticity around any point on the sky.  Primordial $B$-modes can only be produced by gravitational waves and are therefore considered an unambiguous signature of inflationary tensor perturbations.  The energy scale of inflation determines the amplitude of tensor perturbations, and hence the $B$-mode amplitude, so that a detection would fix the inflationary energy scale. The search for primordial $B$-modes is a key effort of observational cosmology \cite{Baumann:2008aj}.
\end{itemize}

The nature of the inflationary epoch is imprinted on the sky in the temperature and polarization anisotropies of the CMB.  In slow-roll inflation, small $\epsilon$ and $\eta$ ensure that the spectra of scalar and tensor fluctuations are nearly scale-invariant.
The shapes of the primordial perturbation spectra
are therefore intimately tied to the inflationary background dynamics as dictated by the shape of the inflaton potential $V(\phi)$.

CMB observations have improved dramatically in the past decade, and near-future experiments will almost certainly continue this trend.
Temperature aniso\-tropies have been measured at the cosmic variance limit over a large range of scales, and experimentalists are now preparing for precision measurements of the polarization of the CMB \cite{Baumann:2008aj}.
A detection of inflationary tensor perturbations via their unique $B$-mode signature would be especially interesting, as their amplitude relates directly to the energy scale of inflation.  This provides a unique opportunity to probe physics at energies near the GUT scale, far out of reach of terrestrial collider experiments.

\subsection{Motivation for Inflation in String Theory}

Besides the intellectual satisfaction of providing a microscopic description of the inflationary process, there are more detailed motivations for studying inflation in the context of string theory.
While inflation is frustratingly effective at making most  signatures of high-energy physics unobservable, e.g.~by exponentially diluting any pre-existing density of GUT-scale relics,  the duration and the details of inflation are nevertheless sensitive to certain aspects of Planck-scale physics.
In the remainder of this review we flesh this out in more detail, but we now briefly preview two examples of the UV sensitivity of inflation that will be central to our discussion.

\subsubsection{Flatness of the Inflaton Potential}

From a top-down perspective the flatness of the inflaton potential in Planck units, as quantified by the slow-roll conditions, Eq.~(\ref{equ:SR}), is a nontrivial constraint.
As we will show in \S\ref{sec:uv},
small (Planck-suppressed) corrections to the potential often induce important corrections to the curvature of the potential,
$\Delta \eta \sim {\cal O}(1)$.
To assess whether inflation can nevertheless occur requires detailed information about Planck-suppressed corrections to the inflaton potential.
This requires either phenomenological assumptions, or preferably microphysical knowledge, about physics at the Planck scale.  String theory equips us to compute such corrections from first principles,
and in \S\ref{sec:case} we give an example of such an analysis for the case of warped D-brane inflation \cite{Baumann:2008kq}.

\subsubsection{Inflationary Gravitational Waves}

As we explain in \S\ref{sec:large}, the UV sensitivity of inflation is particularly strong in models with observable gravitational waves.
A large gravitational wave signal from inflation is associated with a high energy scale
for the inflaton potential and a
super-Planckian
variation of the inflaton field, $\Delta \phi \gg M_{\rm pl}$, between the time when CMB fluctuations were created and the end of inflation.
In \S\ref{sec:large} we explain why theoretical control of the shape of the potential over a super-Planckian range requires certain assumptions about the UV structure of the theory.
Models of large-field inflation are therefore most naturally  studied  in a UV-complete theory, such as string theory.
In \S\ref{sec:tensors} we present the first controlled examples of large-field inflation in string theory \cite{Silverstein:2008sg, McAllister:2008hb}.

Given the exciting possibility of measuring the gravitational wave signature of inflation in the polarization of the CMB, the issue of controlled large-field inflation is of both theoretical and experimental relevance \cite{Baumann:2008aq}.

\subsection{Organization of this Review}

Inflation in string theory is the subject of close to 1,000 papers,
and space considerations prevent us from presenting a truly comprehensive review that summarizes and assesses each important class of models.
(Some representative contributions to the subject include \cite{someimportantworks}; we refer the reader to the reviews \cite{reviews} for a more complete list of references.)
Instead, our goal is an exposition of what we believe to be the primary theme of the subject: the sensitivity of inflation to Planck-scale physics.
As we will explain, this is
the two-way connection by which string theory can clarify inflationary model-building, and cosmological experiments can constrain string theory models.  To illustrate this idea in depth, we focus on two case studies: {\it warped D-brane inflation} and {\it axion monodromy inflation}.
These two scenarios are instructive examples from the general classes of small-field and large-field inflation, respectively.

\section{Inflation in String Theory}
\label{sec:uv}

\subsection{Inflation in Effective Field Theory}
\label{sec:eft}

As a phenomenon in quantum field theory coupled to general relativity, inflation does not appear to be natural. In particular, the set of Lagrangians suitable for inflation is a minute subset of the set of all possible Lagrangians.
Moreover, in wide classes of models, inflation emerges only for rather special initial conditions, e.g.~initial configurations with tiny kinetic energy, in the case of small-field scenarios.  Although one would hope to explore and quantify the naturalness both of inflationary Lagrangians and of inflationary
 initial conditions, the question of initial conditions appears inextricable from the active yet incomplete program of understanding measures in eternal inflation \cite{eternal}.
 (However, see e.g.~\cite{initial} for recent efforts to quantify or to ameliorate the fine-tuning of initial conditions.) In this review we will focus on the question of how (un)natural it is to have a Lagrangian suitable for inflation.

For a single inflaton field with a canonical kinetic term, the necessary conditions for inflation can be stated in terms of the inflaton potential.
 Inflation requires a potential that is quite flat in Planck units (see Eq.~(\ref{equ:SR})), and as we now argue, this condition is sensitive to Planck-scale physics.  Let us recall that the presence of some form of new physics at the Planck scale is required in order to render graviton-graviton scattering sensible, just as unitarity of $W$-$W$ scattering requires new physics at the ${\rm TeV}$ scale.  Although we know that new degrees of freedom must emerge, we cannot say whether the physics of the Planck scale is a finite theory of quantum gravity, such as string theory, or is instead simply an effective theory for some unimagined physics at yet higher scales.  However, the structure of the Planck-scale theory has meaningful -- and, in very favorable cases, testable -- consequences for the form of the inflaton potential.

As usual, the effects of high-scale physics above some cutoff $\Lambda$ are efficiently
described by the coefficients of operators in the low-energy effective theory.  Integrating out particles of mass $M \ge \Lambda$ gives rise to operators of the form
\begin{equation}
\label{equ:ODelta}
\frac{{\cal{O}}_{\delta}}{M^{\delta-4}}\, ,
\end{equation}
where $\delta$ denotes the mass dimension of the operator.

Sensitivity to such operators is commonplace in particle physics: for example, bounds on flavor-changing processes place limits on physics above the ${\rm TeV}$ scale,
and lower bounds on the proton lifetime even allow us to constrain GUT-scale operators that would mediate proton decay.
However, particle physics considerations alone do not often reach beyond operators of dimension $\delta=6$, nor go beyond $M\sim M_{\rm GUT}$.  (Scenarios of gravity-mediated supersymmetry breaking are one exception.)  Equivalently, Planck-scale processes, and operators of very high dimension, are irrelevant for most of particle physics: they decouple from low-energy phenomena.

In inflation, however, the flatness of the potential in Planck units introduces sensitivity
to  $\delta\le 6$ {\it Planck-suppressed} operators, such as
\begin{equation}
\label{equ:O6}
\frac{{\cal{O}}_{6}}{M_{\rm pl}^{2}}\, .
\end{equation}
As we explain in \S\ref{sec:eta}, an understanding of such operators is required to address the smallness of the eta parameter, i.e.~to ensure that the theory supports at least 60~$e$-folds of inflationary expansion.  This sensitivity to dimension-six Planck-suppressed operators is therefore common to all models of inflation.

For large-field models of inflation the UV sensitivity of the inflaton action is dramatically enhanced.
As we discuss in \S\ref{sec:large}, in this important class of inflationary models the potential becomes sensitive to an {\it infinite} series of operators  of arbitrary dimension.

\subsection{The Eta Problem}
\label{sec:eta}

In the absence of any specific symmetries protecting the inflaton potential,
contributions to the Lagrangian of the general form
\beq
\frac{{\cal O}_6}{M_{\rm pl}^2}\, = \, \frac{{\cal O}_4}{M_{\rm pl}^2}\, \phi^2\,
\eeq
are allowed.
If the dimension-four operator ${\cal O}_4$ has a vacuum expectation value
(vev) comparable to the inflationary energy density, $\langle{\cal O}_4 \rangle \sim V$, then this term corrects the inflaton mass by order $H$, or equivalently corrects the eta parameter by order one, leading
to an important problem for inflationary model-building.
Let us reiterate that contributions of this form may be thought of as arising from integrating out Planck-scale degrees of freedom.
In this section we discuss this so-called {\it eta problem} in effective field theory, \S\ref{sec:higgs}, and illustrate the problem in a supergravity example, \S\ref{sec:sugraETA}.

\subsubsection{Radiative Instability of the Inflaton Mass}
\label{sec:higgs}

In a generic effective theory with cutoff $\Lambda$, the mass of a scalar field runs to the cutoff scale unless it is protected by some symmetry.
Since the cutoff for an effective theory of inflation is at least the Hubble scale, $\Lambda \ge H$,
this implies that a small inflaton mass ($m_\phi \ll H$) is radiatively unstable.
Equivalently, the eta parameter receives radiative corrections,
\beq
\Delta\eta = \frac{\Delta m_\phi^2}{3 H^2} \ge 1\, ,
\eeq
preventing prolonged inflation.

The difficulty here is analogous to the Higgs hierarchy problem, but supersymmetry does not suffice to stabilize the inflaton mass: the inflationary energy necessarily breaks supersymmetry, and the resulting splittings in supermultiplets are of order $H$, so that supersymmetry does not protect
a small inflaton mass $m_\phi \ll H$.

In \S\ref{sec:tensors} we discuss the natural proposal to protect the inflaton potential via a shift symmetry $\phi\to\phi +const.$, which is equivalent to identifying the inflaton with
a pseudo-Nambu-Goldstone-boson.
In the absence of such a symmetry the eta problem seems to imply the necessity of fine-tuning the inflationary action in order to get inflation.

\subsubsection{Supergravity Example}
\label{sec:sugraETA}

An important instance of the eta problem arises in locally-supersymmetric theories, i.e.~in supergravity \cite{Copeland:1994vg}.
This case is relevant for many string theory models of inflation because four-dimensional supergravity is the low-energy effective theory of supersymmetric string compactifications.

In ${\cal N} = 1$ supergravity, a key term in the scalar potential is the F-term potential,
\beq
\label{equ:Fterm}
 V_F = e^{K/M_{\rm pl}^2} \left[ K^{\varphi \bar \varphi} D_{\varphi} W \overline{D_{\varphi} W} - \frac{3}{M_{\rm pl}^2} |W|^2 \right]\, ,
\eeq
where $K(\varphi, \bar \varphi)$ and $W(\varphi)$ are the K\"ahler potential and the superpotential, respectively; $\varphi$ is a complex scalar field which is taken to be the inflaton; and
we have defined $D_{\varphi}W \equiv \partial_{\varphi} W + M_{\rm pl}^{-2}(\partial_{\varphi} K) W$.  For simplicity of presentation, we have assumed that there are no other light degrees of freedom, but generalizing our expressions to include other fields is straightforward.

The K\"ahler potential determines
the inflaton kinetic term, $-K_{,\varphi \bar \varphi} \, \partial \varphi \partial \bar \varphi$, while the superpotential determines the interactions.
To derive the inflaton mass,
we expand $K$ around some chosen origin, which we denote by $\varphi \equiv 0$ without loss of generality, i.e.~$K(\varphi, \bar \varphi) = K_0 + \left. K_{,\varphi \bar \varphi}\right|_0 \, \varphi \bar \varphi + \cdots\ $. The inflationary Lagrangian then becomes
\begin{eqnarray}
{\cal L} &\approx& - K_{,\varphi \bar \varphi}\, \partial \varphi  \partial \bar \varphi -  V_0 \Bigl(1+ \left. K_{,\varphi \bar \varphi}\right|_0 \frac{\varphi \bar \varphi}{M_{\rm pl}^2} + \dots \Bigr) \\
&\equiv& -  \partial \phi \partial \bar\phi - V_0 \Bigl(1+  \frac{\phi \bar \phi}{M_{\rm pl}^2} \Bigr) + \dots\, ,
\end{eqnarray}
where we have defined the canonical inflaton field $\phi\bar\phi \approx  \left. K_{\varphi \bar \varphi}\right|_0 \varphi \bar \varphi$ and $V_0 \equiv \left. V_F\right|_{\varphi =0}$.
We have retained the leading correction to the potential originating in the expansion of $e^{K/M_{\rm pl}^2}$ in Eq.~(\ref{equ:Fterm}), which could plausibly be called a universal correction in F-term scenarios. The omitted terms, some of which \
can be of the same order as the terms we keep, arise from expanding $\left[ K^{\varphi \bar \varphi} D_{\varphi} W \overline{D_{\varphi} W} - \frac{3}{M_{\rm pl}^2} |W|^2 \right]$ in Eq.~(\ref{equ:Fterm}) and clearly depend on the model-dependent structure of the K\"ahler potential and the superpotential.

The result is of the form of Eq.~(\ref{equ:O6}) with
\beq
{\cal O}_6 =   V_0\, \phi\bar\phi
\eeq
and implies a large model-independent contribution to the eta parameter
\beq
\label{equ:sugraETA}
\Delta \eta = 1\, ,
\eeq
as well as a model-dependent contribution which is  typically of the same order.  It is therefore clear that in an inflationary scenario driven by an F-term potential, eta will generically be of order unity.

Under what circumstances can inflation still occur, in a model based on a supersymmetric Lagrangian?  One obvious possibility is that the model-dependent contributions to eta approximately cancel the model-independent contribution, so that the smallness of the inflaton mass is a result of fine-tuning.  In the case study of \S3 we will provide a concrete example in which the structure of all relevant contributions to eta can be computed, so that one can sensibly pursue such a fine-tuning argument.

Clearly, it would be far more satisfying to exhibit a mechanism that {\it removes} the eta problem by ensuring that $\Delta \eta \ll 1$.  This requires either that the F-term potential is negligible, or that the inflaton does not appear in the F-term potential.  The first case does not often arise, because F-term potentials play an important role in presently-understood models
for stabilization of the compact dimensions of string theory \cite{Kachru:2003aw, Douglas:2006es}.
However, in \S\ref{sec:tensors} we will present a scenario in which the inflaton is an axion and does not appear in the K\"ahler potential,
or in the F-term potential, to any order in perturbation theory.
This evades the particular incarnation of the eta problem that we have described above.

\subsection{From String Compactifications to the Inflaton Action}

\subsubsection{Elements of String Compactifications}

It is a famous fact that the quantum theory of strings is naturally defined in more than four spacetime dimensions, with four-dimensional physics emerging upon compactification of the additional spatial dimensions.  For concreteness, we will focus on compactifications of the critical ten-dimensional type IIB string theory on six-dimensional Calabi-Yau spaces (to be precise, our compactifications will only be conformal to spaces that are well-approximated by orientifolds of Calabi-Yau manifolds, but we will not need this fine point.)

The vast number of distinct compactifications in this class are distinguished by their topology, geometry, and discrete data such as quantized fluxes and wrapped D-branes.
A central task in string theory model-building is to understand in detail how the ten-dimensional sources determine the four-dimensional effective theory.
If we denote the ten-dimensional compactification data by ${\cal C}$, the procedure in question may be written schematically as
\beq
\label{equ:S10S4}
{\cal S}_{10}[{\cal C}] \quad \to \quad  {\cal S}_4\, .
\eeq
Distinct compactification data ${\cal C}$ give rise to a multitude of four-dimensional effective theories ${\cal S}_4$ with varied field content, kinetic terms, scalar potentials, and symmetry properties.
By understanding the space of possible data ${\cal C}$ and the nature of the map in Eq.~(\ref{equ:S10S4}), we can hope to identify, and perhaps even classify, compactifications that give rise to interesting four-dimensional physics.

\subsubsection{The Effective Inflaton Action}

For our purposes, the most important degrees of freedom of the effective theory are four-dimensional scalar fields.  Scalar fields known as {\it moduli} arise from deformations of the compactification manifold, typically numbering in the hundreds for the Calabi-Yau spaces under consideration, and from the positions, orientations, and gauge field configurations of any D-branes.  From given compactification data one can compute the kinetic terms and scalar potentials of the moduli; in turn, the expectation values of the
moduli determine the parameters of the four-dimensional effective theory.  In the presence of generic ten-dimensional sources of stress-energy, such as D-branes and quantized fluxes, there is an energy cost for deforming the compactification, and many (though not always all) of the moduli fields become massive \cite{Douglas:2006es}.

It is useful to divide the scalar fields arising in ${\cal S}_4$ into a set of light fields $\phi,\psi$ with masses below the Hubble scale ($m_\phi, m_\psi \ll H$) and a set of heavy fields $\chi$ with masses much greater than the Hubble scale ($m_\chi \gg H$).  Here one of the light fields, denoted $\phi$, has been identified as the inflaton candidate.

To understand whether successful inflation can occur, one must understand all the scalar fields, both heavy and light.  First, sufficiently massive moduli fields are effectively frozen during inflation, and one should integrate them out to obtain an effective action for the light fields only,
\beq
{\cal S}_4(\phi, \psi, \chi) \quad \to \quad {\cal S}_{4, \rm eff}(\phi,\psi)\, .
\eeq
Integrating out these heavy modes generically induces contributions to the potential of the putative inflaton: that is, {\it moduli stabilization contributes to the eta problem}.
This is completely analogous to the appearance of corrections from higher-dimension operators in our
discussion of
effective field theory in \S\ref{sec:eft}.

Next, if scalar fields in addition to the inflaton are light during inflation, they typically have important effects on the dynamics, and one should study the evolution of all fields $\psi$ with masses $m_\psi \ll H$.  Moreover, even if the resulting multi-field inflationary dynamics is suitable, light degrees of freedom can create problems for late-time cosmology.  Light scalars absorb energy during inflation and, if they persist after inflation, they can release this energy during or after Big Bang nucleosynthesis, spoiling the successful predictions of the light element abundances.  Moreover, light moduli would be problematic in the present universe, as they mediate fifth forces of gravitational strength. To avoid these late-time problems, it suffices to  ensure that $m_\psi \gg 30 ~\rm{TeV}$, as in this case the moduli decay before Big Bang nucleosynthesis.  A simplifying assumption that is occasionally invoked is that all fields aside from the inflaton should have $m \gg H$, but this is not required on physical grounds: it serves only to arrange that the effective theory during inflation has only a single degree of freedom.

\section{Case Study of Small-Field Inflation: Warped D-brane Inflation}
\label{sec:case}

In string theory models of inflation the operators contributing to the inflaton potential can be enumerated, and in principle even their coefficients can be computed in terms of given compactification data.
To illustrate these issues, it is useful to examine a concrete model in detail.
In the following we therefore present a case study of a comparatively well-understood model of small-field inflation, {\it{warped D-brane inflation}}.

\subsection{D3-branes in Warped Throat Geometries}

\begin{figure}[h!]
\centerline{\psfig{figure=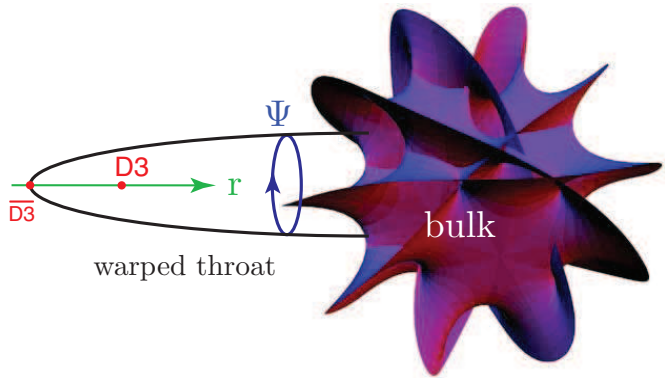,width=0.65\textwidth}}
\vspace{-0.5cm}
\caption{D3-brane inflation in a warped throat geometry.  The D3-branes are spacetime-filling in four dimensions and therefore pointlike in the extra dimensions. The circle stands for the base manifold $X_5$ with angular coordinates $\Psi$. The brane moves in the radial direction $r$. At $r_{\rm UV}$ the throat attaches to a compact Calabi-Yau space. Anti-D3-branes minimize their energy at the tip of the throat, $r_{\rm IR}$.}
\label{fig:throat}
\end{figure}

In this scenario inflation is driven by the motion of a D3-brane in a warped throat region of a stabilized compact space \cite{Kachru:2003sx}.
To preserve four-dimensional Lorentz (or de Sitter) invariance, the D3-brane fills our four-dimensional spacetime and is pointlike in the extra dimensions  (see Figure \ref{fig:throat}).
The global compactification is assumed to be a warped product of four-dimensional spacetime (with metric $g_{\mu \nu}$) and a conformally-Calabi-Yau space,
\beq
\label{equ:metric}
\d s^2 = e^{2A(y)} g_{\mu \nu} \d x^\mu \d x^\nu + e^{-2A(y)} g_{mn} \d y^m \d y^n \, ,
\eeq
with $g_{mn}$ a Calabi-Yau metric that can be approximated in some region by a cone over a five-dimensional Einstein manifold $X_5$,
\beq
g_{mn} \d y^m \d y^n = \d r^2 +r^2 \d s_{X_5}^2\, .
\eeq
A canonical example of such a throat region is the Klebanov-Strassler (KS) geometry \cite{KS}, for which $X_5$ is the $\Bigl(SU(2)\times SU(2)\Bigr)/U(1)$ coset space $T^{1,1}$, and the would-be conical singularity at the tip of the throat, $r=0$, is smoothed by the presence of appropriate fluxes.
The tip of the throat is therefore located at a finite radial coordinate $r_{\rm IR}$, while at $r=r_{\rm UV}$ the throat is glued into an unwarped bulk geometry.
In the relevant regime $r_{\rm IR} \ll r < r_{\rm UV}$ the warp factor may be written as \cite{KT}
\beq
\label{equ:warp}
e^{-4A(r)} = \frac{R^4}{r^4} \ln \frac{r}{r_{\rm IR}}\, , \qquad R^4 \equiv \frac{81}{8} (g_s M \alpha')^2\, ,
\eeq
where
\beq
\ln \frac{r_{\rm UV}}{r_{\rm IR}} \approx \frac{2\pi K}{3 g_s M}\, .
\eeq
Here, $M$ and $K$ are integers specifying the flux background \cite{KS,GKP}.

Warping sourced by fluxes is commonplace in modern compactifications, and there has been much progress in understanding the stabilization of the moduli of such a compactification \cite{Douglas:2006es}.
Positing a stabilized compactification containing a KS throat therefore seems reasonable given present knowledge.

Inflation proceeds as a D3-brane moves radially inward in the throat region, towards an anti-D3-brane that is naturally situated at the tip of the throat.  The inflaton kinetic term is determined by the Dirac-Born-Infeld (DBI) action for a probe D3-brane, and leads to an identification of the canonical inflaton field with a multiple of the radial coordinate, $\phi^2 \equiv T_3 r^2$. Here, $T_3 \equiv \left[(2\pi)^3 g_s \alpha'^2 \right]^{-1}$ is the D3-brane tension, with $g_s$ the string coupling and $2\pi\alpha'$ the inverse string tension.  The exit from inflation occurs when open strings stretched between the approaching pair become tachyonic and condense, annihilating the branes.

In this simplified picture, inflation is driven by the extremely weak (warping-suppressed) Coulomb interaction of the brane-antibrane pair \cite{Kachru:2003sx}. The true story, however, is more complex, as moduli stabilization introduces new terms in the inflaton potential which typically overwhelm the Coulomb term and drive more complicated dynamics \cite{Kachru:2003sx,Baumann:2006th, Baumann:2007np, Baumann:2007ah, Krause:2007jk, Baumann:2008kq}. This pattern is precisely what we anticipated in our effective field theory discussion: integrating out moduli fields can be expected to induce important corrections to the potential.

\subsection{The D3-brane Potential}
\label{sec:gravity}

An important correction induced by moduli stabilization is the inflaton mass term arising from the supergravity F-term potential,
\S\ref{sec:sugraETA}.  In a vacuum stabilized by an F-term potential, i.e.~by superpotential terms involving the moduli,
one finds the mass term $H^2_0 \phi^2 = \frac{1}{3} V_0(\phi_\star) \frac{\phi^2}{M_{\rm pl}^2}$  \cite{Kachru:2003sx},
where $\phi_\star$ is an arbitrary reference value for the inflaton field and the parameter $H_0$ should not be confused with the present-day Hubble constant.

However, one expects additional contributions to the potential from a variety of other sources, such as additional effects in the compactification that break supersymmetry \cite{Baumann:2008kq}.  Let us define $\Delta V(\phi)$ to encapsulate all contributions to the potential aside from the Coulomb interaction $V_{0}(\phi)$ and the mass term $H_0^2 \phi^2$; then the total potential and the associated contributions to the eta parameter may be written as
\begin{eqnarray}
\label{equ:Vphi}
V(\phi) &=& V_{0}(\phi) \ \, + \ H^2_0 \phi^2 \ + \ \Delta V(\phi) \\
\eta(\phi) &=& \quad \eta_0 \quad +  \quad\frac{2}{3} \ \ \ \ + \  \Delta \eta(\phi) \quad = \quad  ?
\end{eqnarray}
where $\eta_0\ll 1$ because the
Coulomb interaction is very weak.
(More generally, $V_0(\phi)$ can be {\it defined} to be all terms in $V(\phi)$ with negligible contributions to $\eta$. Besides the brane-antibrane Coulomb interaction, this can include any other sources of nearly-constant energy, e.g.~bulk contributions to the cosmological constant.)

Clearly, $\eta$ can only be small if $\Delta V$ can cancel the mass term in Eq.~(\ref{equ:Vphi}).  We must therefore enumerate all relevant contributions to $\Delta V$, and attempt to understand the circumstances under which an approximate cancellation can occur.  Note that identifying a subset of contributions to $\Delta V$ while remaining ignorant of others is insufficient.

Warped D-brane inflation
has received a significant amount of theoretical attention in part because of its high degree of computability.  Quite generally, if we had access to the full data of an explicit, stabilized compactification with small curvatures and weak string coupling, we would in principle be able to compute the potential of a D-brane inflaton to any desired accuracy, by performing a careful dimensional reduction.  This is not possible at present for a generic compact Calabi-Yau, for two reasons: for general Calabi-Yau spaces hardly any metric data is available, and examples with entirely explicit moduli stabilization are rare.

However, a sufficiently long throat is well-approximated by a {\it noncompact} throat geometry (i.e., a throat of infinite length), for which the Calabi-Yau metric can often be found,
as in the important example of the Klebanov-Strassler solution \cite{KS}, which is entirely explicit and everywhere smooth.  Having complete metric data greatly facilitates the study of probe D-brane dynamics, at least at the level of an unstabilized compactification.
Furthermore, we will now explain how the effects of moduli stabilization and of the finite length of the throat can be incorporated systematically.  The method involves examining perturbations to the supergravity solution that
describes
the throat in which the D3-brane moves.  For concreteness we will work with the example of a KS throat, but the method is far more general.
Our treatment will allow us to
give explicit expressions for the correction terms $\Delta V$ in Eq.~(\ref{equ:Vphi}), and hence to extract the characteristics of inflation in the presence of moduli stabilization.

\subsection{Supergravity Analysis of the D3-brane Potential}
\label{sec:SG}

\subsubsection{Perturbations to the Geometry}
\label{sec:gravitypert}

Above we gave the explicit solution for the noncompact warped throat region.
We now describe a
systematic way to estimate the leading corrections to the throat solution as perturbations to the geometry and fluxes at large $r$ (near $r_{\rm UV}$ in Figure \ref{fig:throat}). We then consider the effect of these perturbations on
the potential for a D3-brane at a location well inside the throat.

Before we can explain the idea underlying this approach, we need a few facts about the coupling of D3-branes to the background fields of type IIB supergravity,
which is
the low-energy limit of type IIB string theory.
Type IIB supergravity contains a metric, which in the background takes the form of Eq.~(\ref{equ:metric}), as well as various $p$-form fields.
Importantly, the
four-dimensional potential $V$ as a function of the D3-brane position in the extra dimensions is only affected by a very specific combination of background fields:
\beq
V= T_3 \left( e^{4A} - \alpha \right) \equiv T_3\, \Phi_-\, ,
\eeq
where $\alpha$ is the potential for the five-form field
\beq
F_5 = (1+\star_{10}) \left[\d \alpha(y) \wedge \d x^0 \wedge \d x^1 \wedge \d x^2 \wedge \d x^3 \right]\, ,
\eeq
and $\star_{10}$ is the ten-dimensional Hodge star.
The D3-brane potential is therefore dictated by the profile of
$\Phi_- \equiv e^{4A} - \alpha$.
Furthermore, $\Phi_-$ vanishes in the unperturbed KS throat, and more generally in the class of flux compactifications considered by Giddings, Kachru, and Polchinski (GKP) \cite{GKP}.  Finally, the Einstein equations and five-form Bianchi identity imply that perturbations of $\Phi_-$ around such backgrounds satisfy, at the linear level, the six-dimensional Laplace equation \cite{GKP,Baumann:2008kq},
\beq
\label{equ:Laplace}
\nabla^2 \Phi_- = 0\, .
\eeq

In a general compactification, little is known about the solution for $\Phi_{-}$, and one can therefore draw few general conclusions about the D3-brane potential.  However, in a noncompact throat geometry, one can often solve the Laplace equation.  Moreover, the potential for a D3-brane in a throat that is glued into an arbitrary compact Calabi-Yau must be expressible in terms of a solution for $\Phi_{-}$ in the throat, i.e.~in terms of a superposition of harmonic functions.

As this is the crucial idea, let us repeat: no matter how complicated the Calabi-Yau to which the throat is attached, it must be possible to express the D3-brane potential via some $\Phi_{-}$ profile in the throat, as $\Phi_{-}$ is the only supergravity field that sources a D3-brane potential.  Moreover, for a sufficiently long but finite throat, the $\Phi_{-}$ profile is given to an excellent approximation by a solution of the Laplace equation in the noncompact throat geometry.  (Corrections are expected to be exponentially small when the throat is long.) Thus, the structure of the inflaton potential is dictated by the structure of solutions of the Laplace equation in the noncompact throat.

\subsubsection{Harmonic Analysis}

Let us therefore solve the Laplace equation (\ref{equ:Laplace}) in the KS background.
We denote the eigenfunctions of the angular Laplacian on the base manifold $X_5 = T^{1,1}$
by
$Y_{LM}(\Psi)$
 \cite{Ceresole:1999zs}, where the multi-indices $L\equiv (J_1, J_2, R)$, $M \equiv (m_1, m_2)$ label $SU(2) \times SU(2) \times U(1)$ quantum numbers under the corresponding isometries of $T^{1,1}$.
We may then express the solution of the Laplace equation (\ref{equ:Laplace}) as the following expansion
\beq
\label{equ:Phi}
\Phi_-(r, \Psi) = \sum_{L} f_{L}(\Psi) \left( \frac{r}{r_{\rm UV}}\right)^{\Delta(L)} \, ,
\eeq
where
\beq
f_L(\Psi) \equiv \sum_M \Phi_{LM} Y_{LM}(\Psi) + c.c. \, ,
\eeq
and $\Phi_{LM}$ are constant coefficients.
The quantities $\Delta(L)$ are related to the eigenvalues of the angular Laplacian
\beq
\label{equ:Delta}
\Delta(L) \equiv - 2 + \sqrt{6 [J_1(J_1+1) + J_2(J_2+2) -R^2/8] + 4}\ .
\eeq
To determine the {\it leading} perturbations to the brane potential we are interested in the {\it lowest} eigenvalues, since via Eq.~(\ref{equ:Phi}) these correspond to the perturbations that diminish most slowly towards the tip of the throat.
Incorporating the group-theoretic selection rules that restrict the allowed quantum numbers \cite{Ceresole:1999zs}, one finds that the smallest eigenvalues corresponding to nontrivial  perturbations are
\begin{eqnarray}
\Delta = \frac{3}{2} \quad &&{\rm for} \ \ (J_1, J_2, R) = (1/2, 1/2, 1)\, ,\\
\Delta = 2  \quad &&{\rm for} \ \ (J_1, J_2, R) = (1,0,0), (0,1,0)\, .
\end{eqnarray}

For simplicity, we now assume that a single mode dominates the expansion in Eq.~(\ref{equ:Phi}),
\beq
\label{equ:Phi2}
\Phi_- \approx f_{L}(\Psi) \left( \frac{r}{r_{\rm UV}}\right)^{\Delta(L)} \, .
\eeq
(If more than one angular mode is relevant during inflation, then the dynamics is significantly more complicated than what is described below.)
To isolate the radial dynamics, we first minimize the potential in the angular directions. When the angular coordinates have relaxed to their minima, the potential reduces to an effective single-field potential for the radial direction $r$.
In the single-perturbation case of Eq.~(\ref{equ:Phi2}), the $\Phi_-$ perturbation then always leads to a repulsive force, i.e.~the effect of the perturbation is to push the brane out of the throat.
The proof is straightforward: any non-constant spherical harmonic is orthogonal to the constant ($L=0$) harmonic, and hence any nontrivial harmonic necessarily attains both positive and negative values.  Therefore, there always exists an angular location $\Psi_\star$ where $f_L(\Psi_\star)$ is negative. It follows that at fixed radial location, the D3-brane potential induced by the term in Eq.~(\ref{equ:Phi2}) is minimized at an angular location where the contribution of Eq.~(\ref{equ:Phi2}) to the radial potential is negative. This contribution to the potential is minimized at $r\to \infty$. The potential induced by any individual perturbation of the form of Eq.~(\ref{equ:Phi2}) therefore produces a radially-expulsive force.  This is fortunate, since only a repulsive force allows cancellation with the mass term in Eq.~(\ref{equ:Vphi}) to alleviate the eta problem.

\subsubsection{Phenomenological Implications}

If only one angular mode dominates the UV perturbation of the throat geometry, then the radial D3-brane potential is
\beq
V(\phi) = V_0(\phi) + M_{\rm pl}^2 H_0^2 \left[ \left(\frac{\phi}{M_{\rm pl}}\right)^2  - a_\Delta \left( \frac{\phi}{M_{\rm pl}}\right)^\Delta \right]\, .
\eeq
where
\beq
a_\Delta \equiv c_\Delta \left(\frac{M_{\rm pl}}{\phi_{\rm UV}}\right)^{\Delta} \, , \qquad {\rm and} \qquad \phi \propto r\, .
\eeq
The magnitudes of the coefficients $a_\Delta$, $c_\Delta$ are highly model-dependent and were estimated in Ref.~\cite{Baumann:2008kq}.
The above classification of the leading perturbations to the inflaton potential via the eigenvalues of the angular Laplacian hence leads to two cases with distinct phenomenology:

\begin{enumerate}
\item {\sl Fractional Case}

In a general compactification, the dominant perturbation
corresponds to the smallest possible eigenvalue, $\Delta = \frac{3}{2}$. By the repulsivity argument we just gave, this gives a negative contribution to the potential in Eq.~(\ref{equ:Vphi}), $\Delta V \propto - \phi^{3/2}$. The dynamics during inflation is then governed by the following phenomenological potential
\beq
\label{equ:fractional}
V(\phi) = V_0(\phi) + M_{\rm pl}^2 H^2_0 \left[ \left(\frac{\phi}{M_{\rm pl}} \right)^2 - a_{3/2} \left(\frac{\phi}{M_{\rm pl}} \right)^{3/2} \right]\, .
\eeq
For a potential of this form,
the eta parameter can be fine-tuned to be small locally, near an approximate {\it inflection point}.
This inflection point model is phenomenologically identical to the explicit model of D-brane inflation \cite{Baumann:2006th, Baumann:2007np, Baumann:2007ah, Krause:2007jk} in which a moduli-stabilizing D7-brane stack descends into the throat region while wrapping a suitable four-cycle.

\item {\sl Quadratic Case}

Although the $\Delta = \frac{3}{2}$ perturbation is generically dominant, it may be forbidden by a discrete symmetry, i.e.~by an unbroken global symmetry of the full compact manifold \cite{Baumann:2008kq}.
In this case, the leading correction comes from the $\Delta = 2$ perturbation, $\Delta V \propto - \phi^2$.
The relevant phenomenological model is then
\beq
V(\phi) = V_0(\phi) + \beta H^2_0\, \phi^2\, ,
\eeq
where the parameter $\beta$ allows a nearly continuous tuning of the inflaton mass.
The maximally-tuned case $\beta=0$ was first analyzed in Appendix D of Ref.~\cite{Kachru:2003sx} and the phenomenology for general $\beta$ was discussed in Ref.~\cite{Firouzjahi:2005dh}.
(We should note that in the limit $\beta \ll 1$, corrections from perturbations with $\Delta > 2$ can be important.)
As $\beta \to 1$, the potential becomes steep, but inflation may still occur via the DBI effect \cite{Silverstein:2003hf}. The parameter $\beta$ may even be negative, pushing the brane out of the throat and allowing a realization of  IR DBI inflation \cite{Chen:2005ad}.
\end{enumerate}

The above summarizes the phenomenology of warped D-brane inflation under the simplifying assumption that a single angular mode dominates in the $\Phi_-$ perturbation of Eq.~(\ref{equ:Phi2}).
In general, more than one $L$-mode may be important in Eq.~(\ref{equ:Phi2}). In that case we expect the angular dynamics of the brane to be significantly more complicated, with potentially important consequences for the effective single-field potential.

\subsection{Gauge Theory Interpretation}
\label{sec:GT}

\subsubsection{Gauge/Gravity Duality}

In the previous section we have shown how to compute, in supergravity, the leading contributions to the inflaton potential induced by moduli stabilization and by the coupling of the throat region to the compact space.  We will now present a very instructive dual description of this analysis.

The celebrated AdS/CFT correspondence
\cite{ADSCFT} is a duality between type IIB string theory on Anti-de Sitter (AdS) (or asymptotically, approximately AdS) spaces and conformal field theories (CFTs) on their boundaries.  An important class of dual pairs consists of {\it warped throat geometries} and ${\cal{N}}=1$ {\it supersymmetric field theories}.  The system of interest to us, a D3-brane moving in a warped throat solution of type IIB supergravity, therefore admits a dual description in terms of
an approximate CFT.
The corresponding ${\cal{N}}=1$ supersymmetric gauge theory is approximately conformal over a large range of energy scales, and then eventually confines in the infrared.  The gradual deviations from conformality manifest themselves on the gravity side as logarithmic corrections to the warp factor $e^{2A}$, Eq.~(\ref{equ:warp}).

On the gravity side of the correspondence, we were interested in
non-normal\-izable perturbations of the field $\Phi_-$.
In AdS/CFT,
non-normalizable perturbations of supergravity fields correspond to perturbations of the CFT Lagrange density by irrelevant operators,
\begin{equation}
\label{equ:Odelta}
\Delta {\cal L} = M_{\rm UV}^{4-\delta}~ {\cal O}_{\delta}\, ,
\end{equation}
where $M_{\rm UV}$ is the UV cutoff of the gauge theory and ${\cal O}_\delta$ is an operator of mass dimension $\delta \ge 4$.  One advantage of this dual description is that the contributions to the eta parameter are now manifestly organized in terms of operator perturbations, precisely as in our general effective field theory treatment in \S2, cf.~Eq.~(\ref{equ:ODelta}).

We will now outline how the eta problem may be studied on the gauge theory side of the AdS/CFT correspondence, by classifying irrelevant perturbations to the gauge theory. 
For a more complete description we refer the interested reader to Ref.~\cite{Baumann:2008kq}, while readers less interested in these details may skip to \S\ref{sec:perspective} without loss of continuity.

\subsubsection{Perturbations of the Gauge Theory}

The configuration space of a probe D3-brane in a KS throat corresponds to a portion of the Coulomb branch of the dual CFT (i.e.~to a portion of the gauge theory moduli space in which the expectation values of D3-brane collective coordinates reduce the rank of the nonabelian part of the gauge group, but do not reduce the total rank.) Thus, to understand the potential on this configuration space, we are interested in the potential on the Coulomb branch of the CFT. Such a potential can be generated if the CFT Lagrangian is perturbed by operators composed of the scalar fields that parameterize the Coulomb branch.

In particular, we are interested in perturbations that do not explicitly break supersymmetry and that incorporate the effects of bulk Calabi-Yau fields.
The leading such terms are of the form of Eq.~(\ref{equ:Odelta}), with
\beq
\label{ois}
{\cal O}_\delta \equiv \int d^4 \theta \, X^\dagger X \, {\cal O}_\Delta\, ,
\eeq
where $X$ is a bulk moduli field.
This term, being an integral over superspace, is allowed in a supersymmetric Lagrangian, but will break supersymmetry spontaneously if $X$ obtains an F-term vacuum expectation value.
Notice that ${\cal O}_\delta$  is a composite operator, containing both bulk and CFT fields, with total dimension $\delta = 4+\Delta$.
Such a perturbation yields a term in the D3-brane potential of the form \cite{Baumann:2008kq}
\beq
\label{DV}
\Delta V \propto \phi^\Delta \,.
\eeq
In the following we will
identify the CFT operators ${\cal O}_{\Delta}$ that correspond to perturbations of $\Phi_-$ and hence induce a D3-brane potential; the
operator dimensions $\Delta$ will then dictate the structure of possible terms in the D3-brane potential.

\subsubsection{Classification of Operators}

To enumerate the lowest-dimension contributing operators, we must give a few more details of the structure of the gauge theory. (More background on the gauge theory dual to the KS throat may be found in Ref.~\cite{Herzog:2001xk}.)
The approximate CFT that is dual to the KS throat is an $SU(N+M) \times SU(N)$ gauge theory with bi-fundamental fields $A_i, B_i \, (i,j = 1,2)$.  These fields parameterize the Coulomb branch and, in particular, contain the data specifying the D3-brane position.
The single-trace operators
built out of the fields $A_i, B_i$ and their complex conjugates are labeled by their $SU(2)_A\times SU(2)_B \times U(1)_R$ quantum numbers $(J_1, J_2, R)$; this symmetry group corresponds to the isometries of the base manifold $T^{1,1}$.
Using the AdS/CFT correspondence, the dimensions of these operators are given by Eq.~(\ref{equ:Delta}). In fact, the leading contributions  to the D3-brane potential involve either {{\it chiral}} operators whose dimensions are dictated by their $U(1)_R$ charges, or operators related by supersymmetry to the Noether currents of the global symmetries, and in either case the dimensions are protected and could be computed directly in the gauge theory.  However, this will not be true of the operators that induce subleading corrections.

{\it Chiral operators.}
For $J_1= J_2= R/2$, one has chiral operators of the form
\beq
\label{chiral}
{\cal O}_{\Delta}\ = \ {\rm Tr} \Bigl( A^{(i_1} B_{(j_1} A^{i_2} B_{j_2} \ldots A^{i_R)} B_{j_R)} \Bigr) + c.c.
\eeq
The dimensions of these operators, $\Delta= 3R/2$, are fixed by ${\cal N}=1$ superconformal invariance.
The lowest-dimension such operators are
\beq
\label{equ:chiral}
{\cal O}_{3/2} \ = \ {\rm Tr} \left( A_i B_j\right) + c.c.\, ,
 \eeq
 which have $\{J_1, J_2, R\} = \{\frac{1}{2}, \frac{1}{2}, 1\}$. These chiral operators have $\Delta=3/2$, and in generic situations they contribute the leading term in the inflaton potential via
 Eq.~(\ref{DV}).

{\it Non-chiral operators.}
 There are a number of operators which have the next lowest dimension, $\Delta =2$. For example, there are operators with $\{J_1, J_2, R\} = \{1, 0, 0\}$:
 \beq \label{dimtwo}
{\cal O}_2  \ = \quad {\rm Tr}\left( A_1 \bar A_2\right)\ , \quad {\rm Tr} \left(A_2 \bar A_1\right)\ , \quad {1\over \sqrt 2} {\rm Tr} \left(A_1\bar A_1 - A_2 \bar A_2\right)
 \ ,
 \eeq
and the corresponding $\{J_1, J_2, R\} = \{0, 1, 0\}$ operators made out of the fields $B_j$.
These non-chiral operators are in the same supermultiplets as $SU(2)\times SU(2)$ global symmetry currents, and so their dimension is exactly 2.

\vskip 5pt
\begin{table}[h!]
   \centering
        \caption{\bf AdS/CFT Dictionary for Warped D-brane Inflation.}
        \vspace{0.3cm}
   \begin{tabular}{p{1cm}| p{4.5cm} | p{6.3cm}}
 & \hspace{1cm} {\sl Gravity Side} & \hspace{1.2cm} {\sl Gauge Theory Side} \\
  \hline
  \hline
  $\Delta {\cal L}$ & \hspace{1cm} $\Delta V= T_3\ \Phi_-$ \newline \vskip 1pt \vskip -25pt \hspace{1.cm} $\nabla^2 \Phi_- = 0$  \newline   \vskip 1pt \vskip -20pt \hspace{1cm} $ \Delta V = - \left(\frac{\phi}{\phi_{\rm UV}}\right)^\Delta$& \hspace{1cm} $\Delta {\cal L} = \int d^4 \theta \ X^\dagger X\ {\cal O}_{\Delta}$ \newline  \vskip 1pt \vskip -25pt \hspace{0.5cm}  {${\cal O}_{\Delta} = {\rm Tr}\left(A^{(i_1} B_{(j_1} \dots A^{i_R)} B_{j_R)}\right)$} \newline
\vskip 1pt \vskip -20pt \hspace{1cm} $ \Delta V = - \left(\frac{\phi}{\phi_{\rm UV}}\right)^{\Delta}$
  \\
  \vspace{0.2cm}
  $\Delta$ & \vspace{0.2cm} eigenvalue of Laplacian & \vspace{0.2cm} operator dimension\\
 $\phi$ &  radial location & energy scale\\
 $\phi_{\rm UV}$ &  maximal UV radius  &  UV cutoff \\
  \hline
  \hline
     \end{tabular}
   \label{tab:AdSCFT}
\end{table}

\newpage
In the event that a discrete symmetry preserved by the full string compactification forbids the chiral operators ${\cal O}_{3/2}$, the leading contribution to the inflaton potential comes from ${\cal O}_{2}$.
This operator classification precisely matches our supergravity analysis in \S\ref{sec:SG}, and the correspondence 
is summarized in Table \ref{tab:AdSCFT}.
Finally, note
that the contributing composite operators ${\cal O}_\delta$ have dimensions $11/2$ and $6$, in precisely the range that we argued on general grounds in \S2 could yield order-unity contributions to the eta parameter.

\subsection{Summary and Perspective}
\label{sec:perspective}

In \S\ref{sec:uv} we explained how the eta problem is sensitive to dimension-six Planck suppressed operators.
In effective field theory models of inflation one can of course always {\it assume} a solution to the eta problem by a cancellation of the contributing correction terms; in other words, one can postulate that a flat potential $V(\phi)$ arises after an approximate cancellation among dimension-six Planck-suppressed corrections.
In string theory models of inflation, to follow this path would be to abdicate the opportunity to use Planck-suppressed contributions as a (limited) window onto string theory.  Moreover, once $\phi$ is identified with a physical degree of freedom of a string compactification,  the precise form of the potential is in principle fully specified  by the remaining data of the compactification. (Mixing conjecture into the analysis at this stage would effectively transform a `string-derived' scenario into a `string-inspired' scenario; the latter may be interesting as a cosmological model, but will not contribute to our understanding of string theory.)
Thus, overcoming the eta problem becomes a detailed computational question.  One can in principle compute the full 
potential from first principles, and in practice one can often 
classify corrections to the leading-order potential.

In this section, we have enumerated the leading corrections for warped D-brane inflation and showed that an accidental cancellation (or fine-tuning) allows small eta over a limited range of inflaton values.
This gives a non-trivial existence proof for inflationary solutions in warped throat models with D3-branes.

\section{Large-Field Inflation}
\label{sec:large}

The UV sensitivity of inflation described in \S\ref{sec:uv} is vastly increased in the special case of {\it{ large-field }} models, i.e.~scenarios  in which the inflaton traverses a distance in field space larger than the Planck mass.  This class is particularly interesting because it includes every inflationary model that yields a detectably-large primordial gravitational wave signal \cite{Lyth:1996im}, as we now review.

\subsection{The Lyth Bound}

In single-field slow-roll inflation the power spectrum of tensor fluctuations is
\beq
{P}_{\rm t} = \frac{2}{\pi^2} \left(\frac{H}{M_{\rm pl}}\right)^2\, ,
\eeq
where $H$ is the Hubble expansion rate. During inflation $H$ is approximately constant, the background spacetime is nearly de Sitter and quantum fluctuations in any light field such as the metric scale with $H$.
The power spectrum of scalar fluctuations is
\beq
\label{equ:Ps}
{P}_{\rm s} = \left(\frac{H}{2\pi}\right)^2 \left(\frac{H}{\dot \phi} \right)^2\, .
\eeq
The first factor in Eq.~(\ref{equ:Ps}) represents the power spectrum of the inflaton fluctuations (arising from quantum fluctuations in de Sitter space), while the second factor comes from the conversion of fluctuations of the inflaton into fluctuations of the spatial 3-curvature.
On scales smaller than the physical horizon, spatial curvature fluctuations relate to the observed fluctuations in the matter density and
in the CMB temperature.
The ratio between the tensor and scalar fluctuation amplitudes is
\beq
\label{equ:TS}
r \equiv \frac{{P}_{\rm t}}{{P}_{\rm s}} = 8 \left(\frac{\dot \phi}{H M_{\rm pl}} \right)^2\, .
\eeq
If we define $\d N \equiv H \d t$, then we may use Eq.~(\ref{equ:TS}) to write the field variation between the end of inflation, $N_{\rm end}$, and the time when fluctuations on CMB scales were generated, $N_{\rm cmb}$, as the following integral
\beq
\frac{\Delta \phi}{M_{\rm pl}} = \int_{N_{\rm cmb}}^{N_{\rm end}} \d N \left(\frac{r(N)}{8}\right)^{1/2}\, .
\eeq
Since the tensor-to-scalar ratio $r(N)$ is nearly constant during slow-roll inflation,
one can derive the following important relation, originally due to
Lyth \cite{Lyth:1996im}:
\beq
\frac{\Delta \phi}{M_{\rm pl}} \simeq {\cal O}(1) \left(\frac{r_\star}{0.01}\right)^{1/2}\, ,
\eeq
where $r_\star$ is the value of the tensor-to-scalar ratio on CMB scales, $r_\star \equiv r(N_{\rm cmb})$.
In any model with $r_\star>0.01$ one must therefore ensure that $\epsilon, |\eta| \ll 1$ over a super-Planckian range $\Delta \phi > M_{\rm pl}$.

This result implies two necessary conditions for {\it large-field inflation}:
\begin{enumerate}
\item[i)] an obvious requirement is that large field ranges are {\it kinematically} allowed, i.e.~that the scalar field space (in canonical units) has diameter $> M_{\rm pl}$.
This is nontrivial, as in typical string compactifications many fields are not permitted such large excursions. (D3-brane inflation in warped throats, \S\ref{sec:case}, is one class of examples
where the kinematic requirement for large field ranges cannot be fulfilled \cite{Baumann:2006cd}.)
\item[ii)] the flatness of the inflaton potential needs to be controlled {\it dynamically} over a super-Planckian field range.
We discuss this challenge in effective field theory in \S\ref{sec:EFT} and in string theory in \S\ref{sec:tensors}.
\end{enumerate}

\subsection{Super-Planckian Fields and Flat Potentials}
\label{sec:EFT}

To begin, let us consider super-Planckian field excursions in the context of Wilsonian effective field theory.

\subsubsection{No Shift Symmetry}

In the absence of any special symmetries, the potential in large-field inflation becomes sensitive to
an infinite series of Planck-suppressed operators.
The physical interpretation of these terms is as follows: as the inflaton
expectation value changes, any other fields $\chi$ to which the inflaton couples experience changes in mass, self-coupling, etc.
In particular,
any field coupled with at least gravitational strength to the inflaton experiences significant changes when the inflaton undergoes a super-Planckian excursion.  
These variations of the $\chi$ masses and couplings in turn feed back into changes of the inflaton potential and therefore threaten to spoil the delicate flatness required for inflation.
Note that this applies not just to the light ($m\ll H$) degrees of freedom, but even to fields with masses near the Planck scale: integrating out Planck-scale degrees of freedom generically (i.e., for couplings of order unity) introduces Planck-suppressed operators in the effective action.  For nearly all questions in particle physics,  such operators are negligible, but in inflation they play an important role.

The particular operators which appear are determined, as always, by the symmetries of the low-energy action.  As an example, imposing only the symmetry $\phi \to - \phi$ on the inflaton leads to the following effective action:
\beq
{\cal L}_{\rm eff}(\phi) = -\frac{1}{2}(\partial \phi)^2 - \frac{1}{2} m^2 \phi^2 - \frac{1}{4} \lambda \phi^4 - \sum_{p=1}^\infty \left[ \lambda_p \phi^4  + \nu_p (\partial \phi)^2    \right] \left( \frac{\phi}{M_{\rm pl}} \right)^{2p}+ \dots
\eeq
Unless the UV theory enjoys further symmetries,
one expects that the coefficients $\lambda_p$ and $\nu_p$ are of order unity.
Thus, whenever $\phi$ traverses a distance of order $M_{\rm pl}$ in a direction that is not protected by a suitably powerful symmetry, the effective Lagrangian receives substantial corrections from an infinite series of
higher-dimension operators.
In order to have inflation, the potential should of course be approximately flat over a super-Planckian range.  If this is to arise by accident or by fine-tuning, it requires a conspiracy among infinitely many coefficients, which has been termed `functional fine-tuning'
(compare this to the eta problem, \S\ref{sec:eta}, which only requires tuning of one mass parameter).

\subsubsection{Shift Symmetry}

There is a sensible way to control this infinite series of corrections: one can invoke an approximate symmetry that forbids the inflaton from coupling to other fields in any way that would spoil the structure of the inflaton potential.    Such a shift symmetry,
\begin{equation}
\phi \to \phi + const
\end{equation}
protects the inflaton potential in a natural way.
(Proposals using shift symmetries to protect the potential in string inflation include \cite{shift,Dimopoulos:2005ac,McAllister:2008hb}.)

In the case with a shift symmetry, the action of chaotic inflation \cite{Linde:1983gd}
 \beq
 {\cal L}_{\rm eff}(\phi) =- \frac{1}{2} (\partial \phi)^2 - \lambda_p\, \phi^p\, ,
 \eeq
with small coefficient $\lambda_p$ is technically natural. 
However, because we require that this symmetry protects the inflaton even from couplings to Planck-scale degrees of freedom, it is essential that the symmetry should be approximately respected by the Planck-scale theory -- in other words, the proposed symmetry of the low-energy effective action should admit a UV-completion.  Hence, large-field inflation should be formulated in a theory that has access to information about approximate symmetries at the Planck scale.  Let us remark that in effective field theory in general, UV-completion of an assumed low-energy symmetry is rarely an urgent question.  The present situation is different because we do {\it not} know whether all reasonable effective actions can in fact arise as low-energy limits of string theory, and indeed it has been conjectured that many effective theories do not admit UV-completion in string theory \cite{Vafa:2005ui,Ooguri:2006in, Adams:2006sv}.  Therefore, it is important to verify that any proposed symmetry of Planck-scale physics can be realized in string theory.

To construct  an inflationary model with detectable gravitational waves, we are therefore interested in finding, in string theory, a configuration that has both a large kinematic range, and a potential protected by a shift symmetry that is approximately preserved by the full string theory.

\section{Case Study of Large-Field Inflation: Axion Monodromy}
\label{sec:tensors}

We now turn to our second case study, an example of large-field inflation
in string theory.
As we have just discussed, the particular challenge in these models is the need to control an infinite series of contributions to the inflaton potential, arising from couplings of the inflaton to degrees of freedom with masses near the Planck scale. Direct enumeration and fine-tuning of such terms (as in the small-field example in \S\ref{sec:case}) is manifestly impractical, and it appears essential to develop a symmetry argument controlling or forbidding these terms.

An influential proposal in this direction is Natural Inflation \cite{Freese:1990rb},
in which a pseudo-Nambu-Goldstone boson (i.e., an axion) is the inflaton. At the perturbative level, the axion $a$ enjoys a continuous shift symmetry $a\to a+const $ which is broken by nonperturbative effects to a discrete symmetry $a\to a+2\pi$.
The nonperturbative effects generate a periodic potential
\begin{equation}
V(\phi)=\frac{\Lambda^4}{2} \left[ 1- \cos \left({\phi \over f}\right) \right]+ \ldots\, ,
\end{equation}
where $\Lambda$ is a dynamically-generated scale, $f$ is known as the axion decay constant, $\phi \equiv  a f$, and the omitted terms are higher harmonics.
The reader can easily verify from Eq.~(\ref{equ:SR}) that if the omitted terms are negligible and $f>M_{\rm pl}$, this potential can drive prolonged inflation.

As explained above, an important question, in any proposed effective theory in which a super-Planckian field range is protected by a shift symmetry, is whether this structure can be UV-completed.  We should therefore search in string theory for an axion with decay constant $f>M_{\rm pl}$.

\subsection{Axions in String Theory}

\subsubsection{Axions from $p$-Forms}

Axions are plentiful in string compactifications, arising from $p$-form gauge potentials integrated on $p$-cycles of the compact space. For example, in type IIB string theory, there are axions $b_i =2\pi \int_{\Sigma_i} B$ arising from integrating the Neveu-Schwarz (NS) two-form $B$ over two-cycles $\Sigma_i$, as well as axions $c_i =2\pi
\int_{\Sigma_i} C$ arising from the Ramond-Ramond (RR) two-form $C$.
In the absence of additional ingredients such as fluxes and space-filling wrapped branes, the potential for these axions is classically flat and has a continuous shift symmetry which originates in the gauge invariance of the ten-dimensional action. Instanton effects break this symmetry to a discrete subgroup, $b_i \to b_i + 2\pi$ ($c_i \to c_i + 2\pi$).
This leads to a periodic contribution to the axion potential whose periodicity we will now estimate.
We will find that the axion decay constants are smaller than $M_{\rm pl}$ in known, computable limits of string theory \cite{Banks:2003sx,Svrcek:2006yi}.  Readers less familiar with string compactifications can accept this assertion and skip to \S5.2 without loss of continuity.

\subsubsection{Axion Decay Constants in String Theory}

Let $\omega^i$ be a basis for $H^2(X, Z)$, the space of two-forms on the compact space $X$, with
$\int_{\Sigma_i} \omega^j = \alpha'\delta_{i}^{~j}$.
The NS two-form potential $B$ may be expanded as
\begin{equation}
B= \frac{1}{2\pi} \sum_i b_{i}(x)\, \omega^i\, ,
\end{equation}
with $x$ the four-dimensional spacetime coordinate.
The axion decay constant can be inferred from the normalization of the axion kinetic term, which in this case descends from the ten-dimensional term
\beq
\frac{1}{(2\pi)^7 g_s^2\alpha'^4}\int  \d^{10}x {1\over 2} |\d B|^2\quad \supset
\quad \frac{1}{2} \int \d^4 x \sqrt{-g}\, \gamma^{ij} (\partial^{\mu}b_i\partial_{\mu} b_j)\,
 \, ,
\eeq
where
\beq
\label{equ:gamma}
\gamma^{ij} \equiv \frac{1}{6 (2\pi)^9 g_s^2 \alpha'^4} \int_X \omega^i \wedge \star\, \omega^j\,
\eeq
and $\star$ is the six-dimensional Hodge star.
By performing the integral over the internal space $X$ and diagonalizing the field space metric as $\gamma^{ij} \to f_i^2 \delta_{ij}$, one can extract the axion decay constant $f_i$.

It is too early to draw universal conclusions, but a body of evidence suggests that the resulting axion periodicities are always smaller than $M_{\rm pl}$ in computable limits of string theory \cite{Banks:2003sx,Svrcek:2006yi}. 
As this will be essential for our arguments, we will illustrate this result in a simple example. Suppose that the compactification is isotropic, with typical length-scale $L$ and volume $L^6$.  Then using
\beq \alpha' M_{\rm pl}^{2}= \frac{2}{(2\pi)^7}\frac{L^6}{g_s^2\alpha'^3} \eeq
we find from Eq.~(\ref{equ:gamma}) that
\begin{equation}
f^2\approx M_{\rm pl}^2\, \frac{\alpha'^2}{6(2\pi)^2 L^4}\, .
\end{equation}
In controlled compactifications we require $L \gg \sqrt{\alpha'}$, so that $f \ll M_{\rm pl}$.
Qualitatively similar conclusions apply in much more general configurations \cite{Banks:2003sx,Svrcek:2006yi}.

\subsection{Axion Inflation in String Theory}

The above result would seem to imply that Natural Inflation from a single axion field cannot be realized in known string compactifications: string theory provides many axions, but none of these has a sufficiently large field range.  However, there are at least two reasonable proposals to circumvent this obstacle.

\subsubsection{N-flation}

The first suggestion was that a collective excitation of many hundreds of axions could have an effective field range large enough for inflation
\cite{Dimopoulos:2005ac, Easther:2005zr}. This `N-flation' proposal is a specific example of assisted inflation \cite{Liddle:1998jc}, but, importantly, one in which symmetry helps to protect the axion potential from corrections that would impede inflation. Although promising, this scenario still awaits a proof of principle demonstration, as the presence of a large number of light fields leads to a problematic renormalization of the Newton constant, and hence to an effectively reduced field range.
For recent studies of N-flation see \cite{Kallosh:2007cc, Grimm:2007hs}.

\subsubsection{Axion Monodromy}

We will instead describe  an elementary mechanism, {\it{monodromy}}, which allows inflation to persist through multiple circuits of a single periodic axion field.  A system is said to undergo monodromy if, upon transport around a closed loop in the (naive) configuration space, the system reaches a new configuration.  A spiral staircase is a canonical example: the naive configuration space is described by the angular coordinate, but the system changes upon transport by $2\pi$. (In fact, we will find that this simple model gives an excellent description of the potential in axion monodromy inflation.)
The idea of using monodromy to achieve controlled large-field inflation in string theory was first proposed by Silverstein and Westphal \cite{Silverstein:2008sg}, who discussed a model involving a D4-brane wound inside a nilmanifold.  In this section we will focus instead on the subsequent axion monodromy proposal of Ref.~\cite{McAllister:2008hb}, where a monodromy arises in the four-dimensional potential energy
upon transport around a circle in the field space parameterized by an axion.

Monodromies of this sort are possible in a variety of compactifications, but we will focus on a single concrete example.  Consider
type IIB string theory on a Calabi-Yau orientifold, i.e.~a quotient of a Calabi-Yau manifold by a discrete symmetry that includes worldsheet orientation reversal and a geometric involution.  Specifically, we will suppose that the involution has fixed points and fixed four-cycles, known as O3-planes and O7-planes, respectively.  If in addition the compactification includes a D5-brane that wraps a suitable two-cycle $\Sigma$ and fills spacetime, then the axion $b=2\pi \int_{\Sigma}B$ can exhibit monodromy in the potential energy.  (Similarly, a wrapped NS5-brane
produces monodromy for the axion $c=2\pi \int_{\Sigma}C$.)
In other words, a D5-brane wrapping $\Sigma$ carries a potential energy that is {\it not} a periodic function of the axion,
as the shift symmetry of the axion action is broken by the presence of the wrapped brane; in fact, the potential energy increases without bound as $b$ increases.

In the D5-brane case, the relevant potential comes from the Dirac-Born-Infeld action for the wrapped D-brane,
\begin{eqnarray}
S_{\rm DBI} &=& \frac{1}{(2\pi)^5 g_s \alpha'^3} \int_{{\cal M}_4 \times \Sigma}  \d^6 \xi \, \sqrt{\det(G+B)} \\
&=& \frac{1}{(2\pi)^6 g_s \alpha'^2} \int_{{\cal M}_4} \hskip -6pt \d^4 x\, \sqrt{-g} \, \sqrt{(2\pi)^2\ell_\Sigma^4 + b^2} \label{equ:bb}\, ,
\end{eqnarray}
where $\ell_\Sigma$ is the size of the two-cycle $\Sigma$ in string units.
The brane energy, Eq.~(\ref{equ:bb}), is clearly not invariant under the shift symmetry $b\to b +2\pi$, although this is a symmetry of the corresponding compactification without the wrapped D5-brane.  Thus, the DBI action leads directly to monodromy for $b$.  Moreover, when $b \gg \ell_\Sigma^2$, the potential is asymptotically {\it{linear}} in the canonically-normalized field $\varphi_b \propto b$.

\vskip 6pt
Before we give more details of the effects of compactification on the axion potential,
let us qualitatively summarize the inflationary dynamics in this model.  One begins with a D5-brane wrapping a curve $\Sigma$, upon which $ \int_\Sigma B$  is taken to be large.  In other words, the axion $b$ has a large initial vev.  Inflation proceeds by the reduction of this vev, until finally $\int_\Sigma B = 0$  and the D5-brane is
nearly `empty', i.e.~has little worldvolume flux.  During this process the D5-brane does not move, nor do any of the closed-string moduli shift appreciably.  For small axion vevs, the asymptotically linear potential we have described is inaccurate, and the curvature of the potential becomes non-negligible; see Eq.~(\ref{equ:bb}).
At this stage, the axion begins to oscillate around its origin.  Couplings between the axion and other degrees of freedom, either closed string modes or open string modes, drain energy from the inflaton oscillations.  If a sufficient fraction of this energy is eventually transmitted to visible-sector degrees of freedom -- which may reside, for example, on a stack of D-branes elsewhere in the compactification --  then the hot Big Bang begins.  The details of reheating depend strongly on the form of the couplings between the Standard Model degrees of freedom and the inflaton, and this is an important open question, both in this model and in string inflation more generally.  (For representative work on reheating after string inflation, see \cite{reheating}.)

\subsection{Compactification Considerations}

Having explained the essential idea of axion monodromy inflation, we must still ensure that the proposed inflationary mechanism is compatible with moduli stabilization and can be realized in a consistent compactification.  An immediate concern is whether there are additional contributions to the potential, beyond the linear term identified above, that could have important effects during inflation. As we
have
emphasized throughout this review, one expects that in the absence of a symmetry protecting the inflaton potential, generic corrections due to moduli stabilization will contribute $\Delta\eta\sim {\cal O}(1)$.
It is therefore essential to verify that the continuous shift symmetry which protects the inflaton potential is
preserved to an appropriate degree by the stabilized compactification.
For the special case of moduli stabilization in which nonperturbative effects play a role, ensuring that the shift symmetry is not spoiled can be quite subtle.  We will now explain this point, but readers less interested in the details
can skip to \S\ref{sec:axionS}.

\subsubsection{Axion Shift Symmetries in String Theory}

We first observe that a continuous shift symmetry $b \to b+const$ forbids all non-derivative terms in the effective action for $b$, but does not constrain terms involving only the spacetime derivative $\partial_{\mu} b$.  Therefore, the shift symmetry is unbroken to the extent that all non-derivative terms are constrained to vanish.

We now check this criterion in the example of interest by recalling the classic Dine-Seiberg treatment \cite{Dine:1986vd} of axion shift symmetries in string theory.  Dine and Seiberg proved that to any order in perturbation theory (in the absence of D-branes), the effective action for the axion $b$ can only be a function of 
$\partial_\mu b$, i.e.~$b$ has a shift symmetry. To show this, they observed that the zero-momentum coupling of $b$ (corresponding to non-derivative terms) is a total derivative on the worldsheet, and hence vanishes when the worldsheet has no boundary and wraps a topologically trivial cycle in spacetime.

Their argument proceeds as follows.  The two-form $B$ couples to the worldsheet as \cite{Polchinski:1998rq}
\beq \label{equ:dine}
                 \frac{i}{2\pi \alpha'}   \int_{\Sigma} \d^2\xi\, \epsilon^{\alpha\beta}\partial_{\alpha}X^{\mu}\partial_{\beta}X^{\nu}B_{\mu\nu}(X)   \, ,
\eeq
where $X^{\mu}$ are the spacetime coordinates and $\xi^{\alpha}$ are two-dimensional string worldsheet coordinates; i.e.~Eq.~(\ref{equ:dine}) is the pullback of $B$ onto the worldsheet.  If $B$ is imagined to be a constant in spacetime, then the above coupling
is a total derivative on the worldsheet.  Equivalently, the {\it zero-momentum} portion of the axion effective action in spacetime arises from a total-derivative term on the string worldsheet.  (In general backgrounds, $B$ is not constant, but it is the coupling of the constant portion of $B$ that governs zero-momentum  terms in the four-dimensional effective action.)

Hence, if the string worldsheet has no boundary and is topologically trivial, the zero-momentum coupling of the axion $b$ must vanish, and the axion therefore cannot have any non-derivative couplings.  Thus, as long as the worldsheet has no boundary, the axion has no non-derivative couplings to any order in sigma-model perturbation theory (i.e., the perturbation theory of the quantum field theory living on the string worldsheet, whose coupling constant is the inverse spacetime curvature in units of $\alpha'$), because worldsheets wrapping nontrivial  curves in the ten-dimensional spacetime contribute only nonperturbatively, as worldsheet instantons. 

However, closed string worldsheets {\it can} develop boundaries in the presence of D-branes, on which the strings can break and end.  Therefore, in a compactification without D-branes, the shift symmetry of $b$ is preserved to all orders in perturbation theory, while in a compactification containing D-branes, the shift symmetry can be violated.

\subsubsection{The Eta Problem for $b$}

In the present setting, we have deliberately invoked D5-branes, in order to produce a monodromy in the potential.  However, provided that this potential, which we identified with the inflaton potential, is the {\it leading} effect breaking the shift symmetry, the  resulting structure is technically natural.

Although this sounds promising, in the case of the $b$ axion there is in fact an additional ingredient which also breaks the axion shift symmetry.  K\"ahler moduli stabilization is accomplished, in a well-studied class of models \cite{Kachru:2003aw, Douglas:2006es}, by the inclusion of nonperturbative effects, e.g.~from Euclidean D3-branes (D3-brane instantons).  Such effects can circumvent the Dine-Seiberg argument given above, because Euclidean D-branes are nonperturbative effects and provide boundaries for string worldsheets.

We will now sketch the specific difficulty presented by Euclidean D3-branes, referring the interested reader to Ref.~\cite{McAllister:2008hb} for details.  Supposing for simplicity that there is only one K\"ahler modulus, $T$,
the superpotential is of the form $W = W_0 + A \exp(-2\pi\, T)$,  where the exponential term is the Euclidean D3-brane contribution.  At the energy scales in question, $W_0$ and $A$ are constants depending on the stabilized values of the complex structure moduli and of the dilaton.  Furthermore, the K\"ahler potential takes the form \cite{Grimm} $K=-3 M_{\rm pl}^2 \ln(T+\bar T - d\, b^2)$, with $d$ a constant depending on the intersection numbers of the compactification and on the stabilized value of the dilaton.  Although a shift of $b$ can be compensated in the K\"ahler potential by a shift of  $T+\bar T$, the superpotential is then not invariant.  Clearly, the continuous shift symmetry is broken by the nonperturbative superpotential term generated by Euclidean D3-branes.  Euclidean D3-branes therefore make important contributions to the potential of $b$, and in fact generate an eta problem.  One can easily verify \cite{McAllister:2008hb} that this is
precisely analogous to the eta problem in D3-brane inflation.

\subsubsection{Flat Potential for $c$}

The situation may seem discouraging, because even a shift symmetry that was valid to all orders in perturbation theory has turned out to be inadequate to protect the inflaton potential!
However, we will now find an even more potent symmetry in the case of the $c$ axion.

Although  the NS axion $b$ and the RR axion $c$ have many shared features, a crucial distinction is that $c$ couples to D1-branes, via the electric coupling
\beq \label{equ:ccouple}
    \int_{\Sigma}C    \, ,
\eeq
but does {\it not} couple directly to D3-branes (or to D3-brane instantons) that carry vanishing D1-brane charge.
Thus, if the moduli are stabilized exclusively by instantons to which $c$ does not couple, even nonperturbative moduli stabilization will not violate the shift symmetry of $c$.  We refer to Ref.~\cite{McAllister:2008hb} for a description of compactifications in which this mechanism is operative.

These considerations suggest the following scenario.  Instead of a wrapped D5-brane introducing a potential for a $b$ axion, we consider a wrapped NS5-brane that provides a potential for a $c$ axion.  Even in the presence of nonperturbative stabilization of the K\"ahler moduli, such an axion can enjoy the protection of a shift symmetry over a super-Planckian range.  The corresponding inflationary scenario is natural in the technical sense.

\subsection{Summary and Perspective}
\label{sec:axionS}

In \S\ref{sec:large} we showed that an observable gravitational wave signal correlates with the inflaton field moving over a super-Planckian distance during inflation.
Effective field theory models of large-field inflation then require a shift symmetry to protect the flatness of the potential over a super-Planckian range.
It has therefore become an important question whether such shift symmetries arise in string theory and can be used to realize large-field inflation.

In this section, we argued that the
first examples of shift symmetries in string theory that protect the potential over a super-Planckian range are becoming available.
We explained the dual role of the monodromy: i) it results in a large kinematic field range $\Delta \phi > M_{\rm pl}$  by allowing a small fundamental domain to be traversed repeatedly, and ii) in combination with the shift symmetry it controls corrections to the potential over a super-Planckian range.  
The shift symmetry, only weakly broken by $V$, controls corrections $\Delta V$ within a fundamental domain, and 
furthermore relates corrections in one fundamental domain to those in any other.
Monodromy therefore effectively reduces a large-field problem to a small-field problem \cite{Silverstein:2008sg}.

Although more work is required to understand these models and the compactifications in which they arise,
monodromy appears to be a robust and rather promising mechanism for realizing large-field inflation, and hence an observable gravitational wave signal, in string theory.

\section{Outlook}

\subsection{Theoretical Prospects}

As we hope this review has illustrated, theoretical progress in recent years has been dramatic. A decade ago, only a few proposals for connecting string theory to cosmology were available, and the problem of stabilizing the moduli had not been addressed. We now have a wide array of inflationary models motivated by string theory, and the best-studied examples among these incorporate some information about moduli stabilization.  Moreover, a few mechanisms for inflation in string theory have been shown to be robust, persisting after full moduli stabilization with all relevant corrections included.

Aside from demonstrating that inflation is possible in string theory, what has been accomplished? In our view  the primary use of explicit models of inflation in string theory is as test cases, or toy models,
for the sensitivity of inflation to quantum gravity. On the theoretical front, these models have underlined the importance of the eta problem in general field theory realizations of inflation; they have led to mechanisms for inflation that might seem unnatural in field theory, but are apparently natural in string theory; and they have sharpened our understanding of the implications of a detection of primordial tensor modes.

It is of course difficult to predict the direction of future theoretical progress, not least because unforeseen fundamental advances in string theory can be expected to enlarge the toolkit of inflationary model-builders.
However, it is safe to anticipate further gradual progress in moduli stabilization, including the appearance of additional explicit examples with all moduli stabilized; entirely explicit models of inflation in such compactifications will undoubtedly follow.
At present, few successful models exist in M-theory or in heterotic string theory (however, see e.g.~\cite{Mtheory} et seq.),
and under mild assumptions, inflation can be shown to be impossible in certain classes of type IIA compactifications \cite{Hertzberg:2007wc, Flauger:2008ad, Caviezel:2008tf}. It would be surprising if it turned out that inflation is much more natural in one weakly-coupled limit of string theory than in the rest, and the present disparity can be attributed in part to the differences among the moduli-stabilizing tools presently available in the various limits. Clearly, it would be useful to understand how inflation can arise in more diverse string vacua.

The inflationary models now available in string theory are subject to stringent theoretical constraints arising from consistency requirements (e.g., tadpole cancellation) and from the need for some degree of computability. In turn, these limitations lead to correlations among the cosmological observables, i.e.~to predictions.  Some of these constraints will undoubtedly disappear as we learn to explore more general string compactifications. However, one can hope that some constraints may remain, so that the set of inflationary effective actions derived from string theory would be a proper subset of the set of inflationary effective actions in  a general quantum field theory.  Establishing such a proposition would require a far more comprehensive understanding of string compactifications than is available at present.

\subsection{Observational Prospects}
\label{sec:obs}

The theoretical aspects of inflation described in this review are interesting largely because they can be {\it tested} experimentally using
present and future cosmological data.  In order to describe this connection, we will very briefly review recent achievements and near-future prospects in observational cosmology.

\subsubsection{Present and Future Observations} Observations of the cosmic microwave background anisotropies, of the distribution of galaxies on the sky, and of the redshift-luminosity relations of type Ia supernovae have transformed cosmology into an exact science.  This has revealed a strange universe filled with 73\% dark energy, 23\% dark matter, and only 4\% baryons.
In addition, we now have a firm qualitative understanding of the formation of large-scale structures, like the cosmic web of galaxies, through the gravitational instability of small primordial fluctuations.
The perturbation spectrum that forms the seeds of these structures is found to be nearly (but not exactly) scale-invariant, nearly Gaussian, and adiabatic \cite{Komatsu:2008hk}, precisely as predicted by the simplest models of inflation \cite{Lyth:1998xn, Baumann:2008aq}.

Future observations will dramatically extend our knowledge of the primordial fluctuations, probe further details of the inflationary paradigm, and allow us to constrain or exclude a considerable fraction of the proposed scenarios for inflation.  The Planck satellite will measure the temperature anisotropies of the CMB with unprecedented accuracy over a large range of scales; in combination with small-scale CMB experiments (e.g.~ACT \cite{ACT} and SPT \cite{SPT}) this will provide crucial information on deviations of the scalar spectrum from scale-invariance, Gaussianity and adiabaticity.

CMB polarization experiments from the ground (e.g.~Clover \cite{Clover}, QUIET \cite{QUIET}, and BICEP \cite{BICEP}) and from balloons (e.g.~EBEX \cite{EBEX} and SPIDER \cite{Spider}) promise to provide the first significant constraints on inflationary tensor perturbations.
A planned CMB polarization satellite (CMBPol \cite{Baumann:2008aq, Baumann:2008aj, EPIC}) would be designed to be sensitive enough to detect $B$-modes down to a tensor-to-scalar ratio of $r=0.01$, thereby including all models of large-field inflation ($\Delta \phi > M_{\rm pl}$).

\subsubsection{UV Physics in the Sky?}

The most dramatic confirmation of inflation would come from a detection of $B$-mode polarization, which would establish the energy scale of inflation and would indicate that the inflaton traversed a super-Planckian distance.
As we have argued in this review, super-Planckian displacements are a key instance in which the inflaton effective action is particularly sensitive to the physics of the Planck scale.  As  a concrete example of  the discriminatory power of tensor perturbations, any detection of primordial gravitational waves would exclude the warped D3-brane inflation scenario of \S3 \cite{Baumann:2006cd}, while an upper bound $r<0.07$  (or a detection with $r\gg 0.07$) would exclude the axion monodromy scenario of \S5 \cite{McAllister:2008hb}.

A further opportunity arises because single-field slow-roll inflation predicts null results for many cosmological observables, as the primordial scalar fluctuations are predicted to be scale-invariant, Gaussian and adiabatic to a high degree.  A detection of non-Gaussianity, isocurvature fluctuations or a large scale-dependence (running) would therefore rule out single-field slow-roll inflation.  Inflationary effective actions that do allow for a significant non-Gaussianity, non-adiabaticity or scale-dependence often require higher-derivative interactions and/or more than one light field, and such actions arise rather naturally in string theory.  Although we have focused in this review on the sensitivity of the inflaton potential to Planck-scale physics, the inflaton kinetic term is equally UV-sensitive, and string theory provides a promising framework for understanding the higher-derivative interactions that can produce significant non-Gaussianity \cite{Silverstein:2003hf,Alishahiha:2004eh}.

Finally, CMB temperature and polarization anisotropies induced by relic cosmic strings or other topological defects provide probes of the physics of the end of inflation or of the post-inflationary era.
Cosmic strings are automatically produced at the end of brane-antibrane inflation \cite{Sarangi, CMP},
and the stability and phenomenological properties of the resulting cosmic string network are determined by the properties of the warped geometry.  Detecting cosmic superstrings via lensing or through their characteristic bursts of gravitational waves is an exciting prospect.

\subsection{Conclusions}

Recent work by many authors has led to the emergence of robust mechanisms for inflation in string theory.  The primary
motivations for these works are the sensitivity of inflationary effective actions to the ultraviolet completion of gravity, and the prospect of empirical tests using precision cosmological data.  String theory models of inflation have now achieved a reasonable level of theoretical control and are genuinely falsifiable by observational data. Indeed, many
string inflation models are already significantly constrained by the current data \cite{stringdata}.
A more difficult question is how cosmological observations might possibly provide evidence in favor of a string theory model of inflation. Present observations and present theoretical considerations do not oblige us to expect an eventual positive result. However, if we are fortunate enough to detect evidence for string theory in the sky, this will most plausibly arise through a distinctive signature that is unnatural, albeit presumably possible, in field theory models.  Perhaps the best hope would be a striking correlation of many observables.

The theoretical community will eagerly await the coming generation of experimental results \cite{Baumann:2008aq}, in the hope of extracting further clues about the physical properties of the early universe.

\vskip 20pt

{\bf Acknowledgements.} 

We are grateful to
Shamit Kachru, Igor Klebanov, Andrei Linde,
Enrico Pajer, Hiranya Peiris, David Poland, Jesse Thaler, and Eva Silverstein for useful discussions and comments on the draft. LM also thanks K.~Narayan, Angel Uranga, and Marco Zagermann for discussions and correspondence on related topics.

DB thanks the theory group at Cornell and Trident Caf\'e for their hospitality while some of the work on this review was performed.  LM thanks the theory groups at Harvard and IMSc, Chennai, as well as the organizers of the 2008 Indian Strings Meeting, for their hospitality while this review was being prepared.

The research of DB is supported by the David and Lucile Packard Foundation and the Alfred P.~Sloan Foundation.  The research of LM is supported by NSF grant PHY-0355005.


\end{document}